\documentstyle[12pt]{article}
  \def\lsim{\raise0.3ex\hbox{$<$\kern-0.75em\raise-1.1ex\hbox{$\sim$}}}
\def\gsim{\raise0.3ex\hbox{$>$\kern-0.75em\raise-1.1ex\hbox{$\sim$}}}
\def\noi{\noindent}
\def\nn{\nonumber}
\def\bea{\begin{eqnarray}}  \def\eea{\end{eqnarray}}
\def\beq{\begin{equation}}   \def\eeq{\end{equation}}

\def\beeq{\begin{eqnarray}} \def\eeeq{\end{eqnarray}}
\newcommand\mysection{\setcounter{equation}{0}\section}
\renewcommand{\theequation}{\thesection.\arabic{equation}}
\newcounter{hran} \renewcommand{\thehran}{\thesection.\arabic{hran}}

\def\bmini{\setcounter{hran}{\value{equation}}
  \refstepcounter{hran}\setcounter{equation}{0}
  \renewcommand{\theequation}{\thehran\alph{equation}}\begin{eqnarray}}

\def\bminiG#1{\setcounter{hran}{\value{equation}}
\refstepcounter{hran}\setcounter{equation}{-1}
\renewcommand{\theequation}{\thehran\alph{equation}}
\refstepcounter{equation}\label{#1}\begin{eqnarray}}

\def\emini{\end{eqnarray}\relax\setcounter{equation}{\value{hran}}\renewcommand{\theequation}{\thesection.\arabic{equation}}}

\begin{document}
\begin{center}
{\large \bf Two-Dimensional Coulomb Systems on a Surface} \\  
{\large \bf of Constant Negative Curvature} \\
\vskip 8 truemm
{\bf B. Jancovici}\footnote{Laboratoire de Physique Th\'eorique et Hautes Energies, Universit\'e
de Paris-Sud, B\^atiment 210, F-91405 Orsay Cedex, France (Laboratoire Associ\'e au Centre National
de la Recherche Scien\-ti\-fi\-que~- URA D0063). E-mail : janco@stat.th.u-psud.fr} {\bf and G.
T\'ellez}\footnote{Laboratoire de Physique, Ecole Normale Sup\'erieure de Lyon, 46 all\'ee
d'Italie, 69364 Lyon Cedex 07, France (Laboratoire Associ\'e au Centre National
de la Recherche Scientifique~- URA 1325). 
E-mail :
gtellez@physique.ens-lyon.fr}
\end{center}

\vskip 5 truemm

\begin{abstract}
We study the equilibrium statistical mechanics of classical two-dimensional Coulomb systems living
on a pseudosphere (an infinite surface of constant negative curvature). The Coulomb potential
created by one point charge exists and goes to zero at infinity. The pressure can be expanded as a
series in integer powers of the density (the virial expansion). The correlation functions have a
thermodynamic limit, and remarkably that limit is the same one for the Coulomb interaction and some
other interaction law. However, special care is needed for defining a thermodynamic limit of the
free energy density. There are sum rules expressing the property of perfect screening. These
generic properties can be checked on the Debye-H\"uckel approximation, and on two exactly solvable
models~: the one-component plasma and the two-component plasma, at some special temperature.  
\end{abstract}

\vskip 5 truemm
\noindent {\bf KEY WORDS :} Pseudosphere ; two-dimensional Coulomb systems ; Coulomb potential ;
virial expansion ; screening ; exactly solvable models. \par
\vskip 1 truecm

\noi LPTHE Orsay 98-01 \par
\noi January 1998 \par
\newpage
\pagestyle{plain}
\mysection{INTRODUCTION}
\hspace*{\parindent} 
How the properties of a system are affected by the curvature of the space in which the system
lives is a question which arises in general relativity. This is an incentive for studying simple
models.

The classical equilibrium statistical mechanics of Coulomb systems (charged particles interacting
by Coulomb's law) has been much studied. Coulomb systems have interesting generic properties,
related to the screening effect. For two-dimensional systems (such as particles in a plane with a
logarithmic interaction), there are exactly solvable models which are useful for checking and
illustrating the generic properties. It might be of some interest to investigate what are the
curvature effects. 

This problem has already been considered for Coulomb systems living on a sphere \cite{1r}-\cite{3r}
or an hypersphere \cite{4r,5r}. One of the motivations for studying such systems is that they are
convenient for numerical simulations, since they are finite without boundaries. However, that very
finiteness has drawbacks from a fundamental point of view, since it is not possible to take the
thermodynamic limit (infinite system) without suppressing the curvature at the same time~; thus one
cannot get information about the effect of curvature on phenomena such as screening or phase
transitions pertaining to infinite systems.

In two dimensions, a surface of constant negative curvature may have the in\-te\-res\-ting feature of
being both curved and infinite. The present paper is about Coulomb systems living on such a surface.
In Section 2, salient properties of that surface are reviewed. The relevant Coulomb potential is
discussed in Section 3. Section 4 is about general features of statistical mechanics on the curved
surface. Screening effects are studied in Section 5. Then, more specific models are considered~: the
Debye-H\"uckel approximation in Section 6, the exactly solvable one-component plasma in Section 7,
the exactly solvable two-component plasma in Section 8.

\mysection{PSEUDOSPHERE} 

\hspace*{\parindent} Let us recall a few properties of the surface of constant negative curvature
called a pseudosphere \cite{6r}. Such a surface is a two-dimensional manifold, the entirety of which
cannot be embedded in three-dimensional Euclidean space. The surface can be represented by a plane
disk of radius $2a$ centered at the origin, the Poincar\'e disk, with a metric such that  

\beq
ds^2 = {dr^2 + r^2 d\varphi^2 \over \left ( 1 - {r^2 \over 4a^2} \right )^2}
\label{2.1e}
\eeq

\noi in terms of the polar coordinates $(r, \varphi )$. The geodesics are circle arcs normal to the
boundary circle of the disk. The geodesic distance from any point inside the disk to the boundary
circle is infinite~: the size of the surface represented by the Poincar\'e disk is infinite. The
Gaussian curvature has the constant value $- 1/a^2$ (our definition of the Gaussian curvature is
such that it is $1/R^2$ for a sphere of radius $R$).

Another convenient set of coordinates is $(\tau , \varphi )$ , with $\tau$ defined through

\beq
\label{2.2e}
r = 2a \tanh {\tau \over 2} 
\eeq

\noi The corresponding expression for the metric is

\beq
\label{2.3e}
ds^2 = a^2 \left ( d \tau^2 + \sinh^2 \tau \ d \varphi^2 \right )
\eeq

\noi The geodesic distance between two points $(\tau , \varphi )$ and $(\tau ' , \varphi ')$ is
given by 

\beq
\label{2.4e}
\cosh {s \over a} = \cosh \tau \cosh \tau ' - \sinh \tau \sinh \tau ' \cos (\varphi - \varphi ')
\eeq

\noi In particular, the geodesic distance to the origin is $a\tau$. 

This representation of a pseudosphere by a Poincar\'e disk is very
similar to the representation of a sphere of radius $R$ by its stereographic projection from its
North pole onto the plane tangent to its South pole \cite{2r}. Here, $R^2$ is replaced by $- a^2$,
the spherical coordinate $\pi - \theta$ is replaced by $\tau$, trigonometric functions are replaced
by hyperbolic ones. An important difference is that the (finite) sphere is represented by the whole
plane, while the (infinite) surface of constant negative curvature is represented by the inside of
the Poincar\'e disk only. 

It should be noted that, for a system of particles living on a surface of constant negative
curvature, having a uniform density $n$ means that the average number of particles in the area
element

\beq
\label{2.5e}
dS = {rdr \ d \varphi \over \left ( 1 - {r^2 \over 4a^2} \right )^2}
\eeq

\noi is $ndS$ and all points are equivalent. However, in the Poincar\'e disk representation, there
will be an apparent non-uniform density $n[1 - (r^2/4a^2)]^{-2}$ which increases up to infinity as
$r$ approaches its upper value $2a$. 

Through an appropriate conformal transformation, the Poincar\'e disk can be mapped onto the upper
half-plane, the Poincar\'e half-plane. This provides another, often used, representation. Here, it
will be more convenient to use the Poincar\'e disk.

\mysection{THE COULOMB POTENTIAL}
\hspace*{\parindent}
The Laplacian is

\bea
\Delta &=& \left ( 1 - {r^2 \over 4a^2} \right )^2 \left [ {\partial^2 \over \partial r^2} + {1
\over r} {\partial \over \partial r} + {1 \over r^2} {\partial^2 \over \partial \varphi^2} \right ]
\nn \\
&=& {1 \over a^2} \left [ {1 \over \sinh \tau } {\partial \over \partial \tau} \sinh \tau {\partial
\over \partial \tau} + {1 \over \sinh^2 \tau} {\partial^2 \over \partial \varphi^2} \right ]
\label{3.1e} \eea

\noindent The Coulomb potential $v(s)$ at a geodesic distance $s$ from a unit point charge is
defined by

\beq
\label{3.2e}
\Delta v(s) = - 2 \pi \ \delta^{(2)}(s)
\eeq

\noi together with the boundary condition that $v(s)$ vanishes at infinity. In the present
paper, unless otherwise specified, $\delta^{(2)}$ is the Dirac distribution on the curved
manifold, i.e. such that

\beq
\int \delta^{(2)} (s) dS = 1
\label{3.3}
\eeq

\noindent where $dS$ is an area element defined with the metric (\ref{2.1e}). The solution of
(\ref{3.2e}) exists and is

\beq
v(s) = - \log \tanh {s \over 2a}
\label{3.3e}
\eeq 

This result should be contrasted with the usual two-dimensional Coulomb potential in a flat space,
$- \log (s/\ell )$, which cannot be made to vanish at infinity, whatever the choice of the constant
length scale $\ell$ might be.

One should also note that, on a sphere, (\ref{3.2e}) has no solution, and one resorts to the choice
\cite{1r} $- \log \sin (s/2R)$ which is the Coulomb potential created by a $+ 1$ point charge
{\it and} a $- 1$ charge uniformly spread on the sphere, or alternatively the choice \cite{7r}
$- \log \tan (s/2R)$ which is the Coulomb potential created by a $+ 1$ point charge
{\it and} a $-1$ point charge at the antipodal point. The difference is, of course, that, on
a sphere, the field lines originating from a single point charge converge into another singularity at
the antipodal point, while, on a pseudosphere, the field lines originating
from a point charge spread away to infinity.

\mysection{STATISTICAL MECHANICS}
\hspace*{\parindent}
We now address the problem of building the classical equilibrium statistical mechanics of an
infinite system of charged particles living on a pseudosphere. Two
particles of charges $q_i$ and $q_j$ interact through the Coulomb potential $- q_i q_j \log \tanh
(s_{ij}/2a)$, where $s_{ij}$ is the geodesic distance between these particles. We shall also
consider the one-component plasma model, made of one species of positive particles of charge $q$
and a uniformly negatively charged background. What are the generic properties to be expected~?

\subsection{Thermodynamic functions}
\hspace*{\parindent}
On a pseudosphere, there are problems with the definition of the
thermodynamic functions. For instance, for defining the free energy density, the standard procedure
would be to compute the free energy $F$ in a domain of finite area $S$ and to take the limit of
$F/S$ as the domain becomes infinite, keeping the particle number density $n$ fixed. In the case of a
flat system, the limit does not depend on the boundary conditions, because the boundary effects
are of order $S^{1/2}$ while the bulk effects are of order $S$. However, in the present case of
a pseudosphere, the length $L$ of the boundary and the area $S$ are of the same
order. Indeed, a ``geodesic disk'' of radius $s$, i.e. the set of points which have a geodesic
distance to some center smaller than $s$, has an area $S = 4 \pi a^2 \sinh^2 (s/2a)$ while the
length of its boundary is $L = 2 \pi a \sinh (s/a)$. As $s \to \infty$, both $S$ and $L$ are of
order $\exp (s/a)$. Therefore, the boundary contributions to $F$ are important, and there is no
unique thermodynamic limit of $F/S$. 

There are nevertheless reasonable ways of defining a free energy density or other thermodynamic
functions. Some of these approaches will be described in the following. Yet, it should be kept
in mind that it is essential to get rid of the boundary effects.

\subsection{The virial expansion}
\hspace*{\parindent}
It is well-known that, for flat conducting Coulomb systems, the pressure cannot be expanded in
integer powers of the density~; the pressure is not an analytical function of the density in a
neighborhood of the origin. For those systems which, at low temperature, go into a dielectric
phase through a Kosterlitz-Thouless transition, the non-analyticity can be used as a signature of
the conducting phase \cite{8r,9r}.

It will now be shown that, on a pseudosphere, things appear to be
different. There are strong indications that the virial expansion of the pressure in powers of the
density exists for Coulomb systems which however will be shown in Sections 6 and 7 to be conductors.

At temperature $T$ (as usual, we set $\beta = 1/k_BT$ with $k_B$ the Boltzmann constant), the
virial expansion of the pressure $p$ with respect to the total number density $n$ would be 

\beq
\label{4.1e}
\beta p = n + \sum_{k=2}^{\infty} B_k \ n^k
\eeq

For simplicity, we shall only discuss in detail the second virial coefficient $B_2$, for two
models. \par \vskip 5 truemm

{\bf 4.2.1 Two-component plasma.}  This is a model made of two species of particles of charges $\pm
q$, interacting through the Coulomb potential (\ref{3.3e}). The second virial coefficient is

\beq
B_2 = - {1 \over 4} \int \left [ e^{-\beta q^2v(s)} + e ^{\beta q^2v(s)} - 2 \right ] dS
\label{4.2e}
\eeq

\noi where we have regrouped the Mayer bonds for like and unlike particles. The area element is

\beq
dS = 2 \pi a \sinh \left ( {s \over a} \right ) ds
\label{4.3e}
\eeq

\noi At large distances $s$, the square bracket in (\ref{4.2e}) behaves like $4(\beta q^2)^2 \exp
(-2s/a)$, and therefore the integral converges in spite of the $\sinh (s/a)$ in $dS$ (some
short-range potential should be added, if necessary, for making the integral convergent at short
distances). This is to be contrasted with what happens for a flat system where $B_2$ diverges. 

We conjecture that a similar analysis would show the higher-order virial coefficients to be also
finite.

When $\beta q^2 < 2$, no short-range potential is needed for making (\ref{4.2e}) convergent. With
$v(s)$ the pure Coulomb potential (\ref{3.3e}), $B_2$ can be explicitly computed. Through the change
of variable $t = \tanh^2 (s/2a)$ and an integration by parts, one obtains

\bea
\label{4.4}
B_2 &=& - {\pi \over 2} \beta q^2 a^2 \int_0^1 {t^{-\beta q^2/2} - t^{\beta q^2/2} \over 1 - t} dt
\nn \\
&=& - {\pi \over 2} \beta q^2 a^2 \left [ \psi \left ( 1 + {\beta q^2 \over 2} \right ) - \psi \left
( 1 - {\beta q^2 \over 2} \right ) \right ] \eea

\noi where $\psi$ is the logarithmic derivative of the gamma function, which has an integral
representation \cite{10r} which provides the second equality in (\ref{4.4}). \par \vskip 5 truemm

{\bf 4.2.2 One-component plasma.} This is a model made of one species of particles of charge $q$ and
a uniform background charged with the opposite sign. The second virial coefficient now is

\beq
B_2 = - {1 \over 2} \int \left [ e^{-\beta q^2v(s)} - 1 + \beta q^2 v(s) \right ] dS
\label{4.4e}
\eeq

\noi where we have added to the Mayer bond the contribution $\beta q^2v (s)$ from the
particle-background and background-background interactions. At large distances, the square bracket
in (\ref{4.4e}) behaves like $2(\beta q^2)^2 \exp (-2s/a)$ and $B_2$ is finite.

The explicit calculation of $B_2$ can be performed through the same change of variable as above $t
= \tanh^2 (s/2a)$, with the result

\bea
\label{4.6}
B_2 &=& - \pi \beta q^2 a^2 \int_0^1 {1 - t^{\beta q^2/2} \over 1 - t} dt \nn \\
&=& - \pi \beta q^2 a^2 \left [ \gamma + \psi \left ( 1 + {\beta q^2 \over 2} \right ) \right ]
\eea

\noi where $\gamma = 0.577 \cdots$ is Euler's constant. \par \vskip 5 truemm

{\bf 4.2.3 A possible definition of the thermodynamic functions.} If indeed the virial expansion
(\ref{4.1e}) exists, it provides, in principle, a definition of the pressure free from boundary
effects. This happens because the thermodynamic limit of the virial coefficients $B_k$ has been taken
before the summation (\ref{4.1e}). The other thermodynamic functions such as the free energy per
particle $f$ or the internal energy per particle $u$ can be obtained through the relations

\beq
p = n^2 {\partial f \over \partial n} \qquad , \qquad u = {\partial (\beta f ) \over \partial
\beta} 
\label{4.7}
\eeq

\noi Of course, in practice, the exact
resummation of the expansion (\ref{4.1e}) will not be feasible in general. 

The low-density
expansion of the excess free energy per particle $f_{exc}$ starts as 

\beq
\beta f_{exc} = B_2 n + \cdots 
\label{4.8}
\eeq

\noi The corresponding expansions of the excess internal energy per particle are 

- for the two-component plasma

\bea
\label{4.9}
u_{exc} &=& - {\pi \over 2} q^2 a^2 \Big [ \psi \left ( 1 + {\beta q^2 \over 2} \right ) - \psi
\left ( 1 - {\beta q^2 \over 2} \right ) + {\beta q^2 \over 2} \psi ' \left ( 1 + {\beta q^2 \over
2} \right ) \nn \\
&+& {\beta q^2 \over 2} \psi ' \left ( 1 - {\beta q^2 \over 2} \right ) \Big ] n +
\cdots
\eea

- for the one-component plasma

\beq
\label{4.10}
u_{exc} = - \pi q^2 a^2 \left [ \gamma + \psi \left ( 1 + {\beta q^2 \over 2} \right ) 
+{\beta q^2 \over 2} \psi ' \left ( 1 + {\beta q^2 \over 2} \right ) \right ] n + \cdots 
 \eeq

\subsection{Correlations}
\hspace*{\parindent}
In contrast with the thermodynamic functions, the $n$-body densities are expected to have
well-defined thermodynamic limits. This is related to the screening properties, to be discussed in
Section~5~: screening makes the $n$-body densities inside the system, far away from the boundary,
insensitive to the boundary conditions.

Screening has a remarkable consequence~: in the thermodynamic limit, the correlation functions are
the same ones in a system with the Coulomb interaction $-\log \tanh (s/2a)$ or when this
interaction is replaced by $-\log \sinh (s/2a)$. This can be shown as follows. 

>From (\ref{2.2e}) and
(\ref{2.4e}), it is found that, in terms of the complex coordinate $z = r \exp (i \varphi )$ in the
Poincar\'e disk, the geodesic distance $s_{ij}$ between two particles located at $z_i$ and $z_j$ is
such that

\beq
\tanh {s_{ij} \over 2a} = \left | {(z_i - z_j)/2a \over 1 - (z_i \bar{z}_j/4a^2)} \right |
\label{4.5e}
\eeq 

\noi and

\beq
\sinh {s_{ij} \over 2a} = {|z_i - z_j|/2a \over \left [ 1 - (r_i/2a)^2 \right ]^{1/2} \left [ 1 -
(r_j/2a)^2 \right ]^{1/2}}
\label{4.6e}
\eeq

\noi For a system of particles of charges $q_i$, with the interaction $-q_i q_j \log \tanh
(s_{ij}/2a)$, the total potential energy can be written

\bea
H_1 &=& - {1 \over 2} \sum_{i\not= j} q_i q_j \log \left | {(z_i - z_j)/2a \over 1 - (z_i
\bar{z}_j/4a^2)} \right | + {1 \over 2} \sum_i q_i^2 \log \left [ 1 - (r_i/2a)^2 \right ] \nn \\
&&- {1 \over 2} \sum_i q_i^2 \log \left [ 1 - (r_i/2a)^2 \right ] 
\label{4.7e} 
\eea

The first two terms in the right-hand side of (\ref{4.7e}) are the potential energy of a system of
particles in a flat disk of radius $2a$ with ideal conductor walls at zero potential (the
second-term is the interaction of each particle with its own image). On the other hand, for a
system of particles with the interaction $-q_i q_j \log \sinh (s_{ij}/2a)$, the total potential
energy can be written as

\beq
H_2 = - {1 \over 2} \sum_{i \not= j} q_i q_j \log {|z_i - z_j| \over 2a} - {1 \over 2} \sum_i q_i^2
\log \left [ 1 - (r_i/2a)^2 \right ]
 \label{4.8e}
\eeq

\noi (the system has been assumed to be neutral, thus $\sum\limits_i q_i = 0$ has been used). The
first term in the right-hand side of (\ref{4.8e}) is the potential energy of a system of particles
in a flat domain with plain hard walls. In both (\ref{4.7e}) and (\ref{4.8e}), the last
term is a same one-body potential associated with the curvature. Comparing (\ref{4.7e}) and
(\ref{4.8e}), we see that the only difference between them is the nature of the walls. Since the
apparent density in the Poincar\'e disk (number of particles in $d^2z$ divided by $d^2z$) becomes
infinite at its boundary (because of the metric (\ref{2.1e})), the apparent screening length will
vanish at the boundary, and the correlation functions inside the Poincar\'e disk will be independent
of the nature of the wall.

Similar consideration hold for a one-component plasma.

In the following, for computing correlation functions, choosing the interaction $- \log \tanh
(s/2a)$ or $- \log \sinh (s/2a)$ will just be a matter of convenience. In some cases, it will be
explicitly checked that both interactions give the same correlation functions. The interaction $-
\log \sinh (s/2a)$ has the advantage of being the analog of $- \log \sin (s/2R)$ on a sphere
\cite{1r}, and that makes possible to use previously developed methods.

Of course, both interactions have the same behavior $- \log (s/2a)$ in the flat system limit $a \to
\infty$. 

\mysection{SCREENING AND SUM RULES}
\hspace*{\parindent}
Screening is a characteristic property of conductors. Internal screening means that, at
equilibrium, a particle of the system is surrounded by a polarization cloud of opposite charge.
External screening means that, at equilibrium, an external charge introduced in the system is
surrounded by a polarization cloud of opposite charge. For a flat system, internal screening and
external screening of an infinitesimal charge, respectively, result in the two Stillinger-Lovett sum
rules \cite{11r} obeyed by the correlation functions. 

In this Section, assuming that screening holds, we derive the generalization of these
Stillinger-Lovett sum rules to the case of a Coulomb system on a pseudosphere. 

\subsection{Internal screening}
\hspace*{\parindent}
Let us consider a system of particles of several species $\alpha$. Each particle of species
$\alpha$ carries a charge $q_{\alpha}$. The number density of species $\alpha$ is $n_{\alpha}$. The
pair correlation functions are $h_{\alpha \beta}(s)$ where $s$ is the geodesic distance between the
two particles. Internal screening is expressed by the sum rule 

\beq
\label{5.1e}
\int \sum_{\beta} n_{\beta} \ q_{\beta} \ h_{\alpha \beta}(s) \ dS = - q_{\alpha}
\eeq

\subsection{External screening}
\hspace*{\parindent}
If an external point charge $Q$ is introduced into the system, it induces a charge density
$\rho_Q(s)$ at a distance $s$ from this external charge. External screening means that

\beq
\label{5.2e}
\int \rho_Q(s) \ dS = - Q
\eeq

In the case when $Q$ is infinitesimal, linear response theory allows to transform (\ref{5.2e}) into
a sum rule involving the correlation functions of the system in the absence of the external charge.
Indeed, if the charge $Q$ is at $z$, its interaction energy with the system is $\widehat{H}_{int} =
Q\widehat{\phi} (z)$ where $\widehat{\phi}(z)$ is the microscopic electric potential created at $z$
by the system and the average induced charge density at $z'$ is 

\beq
\label{5.3}
\rho_Q (z') = - \beta < \widehat{\rho}(z') \widehat{H}_{int}> = - \beta Q <\widehat{\rho}(z')
\widehat{\phi}(z)> \eeq

\noi where $\widehat{\rho}(z')$ is the microscopic charge density at $z'$. Using (\ref{5.2e}), one
obtains the Carnie and Chan sum rule \cite{12r}

\beq
\label{5.4}
\beta \int <\widehat{\rho}(z') \widehat{\phi} (z) > dS' = 1
\eeq

\noi Since $<\widehat{\rho}(z') \widehat{\phi}(z)>$ depends only on the distance $s$ between $z$
and $z'$, (\ref{5.4}) can be rewritten as

\beq
\label{5.5}
2 \pi a^2 \beta \int_0^{\infty} < \widehat{\rho}(0) \widehat{\phi}(\tau )> \sinh \tau \ d\tau = 1
\eeq

\noi and transformed by two successive integrations by parts into 

\beq
\label{5.6}
4 \pi a^2 \beta \int_0^{\infty} \left ( \log \cosh {\tau \over 2} \right ) \left [ {1 \over \sinh
\tau} \ {d \over d \tau} \sinh \tau {d \over d \tau} < \widehat{\rho}(0) \widehat{\phi} (\tau ) >
\right ] \sinh \tau d\tau = 1
  \eeq

\noi where the Laplacian appears. From the Poisson equation,

\beq
\label{5.7}
{1 \over a^2} \ {1 \over \sinh \tau} \ {d \over d \tau} \sinh \tau {d \over d \tau} <
\widehat{\rho}(0) \widehat{\phi} (\tau ) > = - 2 \pi < \widehat{\rho}(0) \widehat{\rho} (\tau ) >
\eeq

\noi Thus, we obtain the sum rule

\beq
\label{5.3e}
4 \pi a^2\beta \int \left ( \log \cosh {s \over 2a} \right ) \rho^{(2)}(s) \ dS = - 1
\eeq

\noindent where $\rho^{(2)}(s) = <\widehat{\rho}(0) \widehat{\rho}(\tau )>$ is the charge pair
correlation function. For a flat two-dimensional system ($a \to \infty$), (\ref{5.3e}) reduces to
the usual second Stillinger-Lovett sum rule

\beq
\label{5.4e}
{\pi \beta \over 2} \int_0^{\infty} r^2 \rho^{(2)}(r) \ 2 \pi r \ dr = - 1
\eeq 

\mysection{DEBYE-H\"UCKEL APPROXIMATION} 
\hspace*{\parindent}
Although the Debye-H\"uckel approximation can be formulated for any many-component plasma, for the
sake of simpler notation we consider only the case of a one-component plasma with number density
$n$ and particle charge $q$ (the background has a charge density $- qn$). The Debye-H\"uckel
approximation is expected to be valid in the weak coupling limit $\beta q^2 \ll 1$. 

\subsection{Correlations}
\hspace*{\parindent}
The average electric potential $\phi (s)$ at distance $s$ from a given particle obeys the Poisson
equation 

\beq
\Delta \phi (s) = - 2 \pi q \left [ nh(s) + \delta^{(2)}(s) \right ]
\label{6.1}
\eeq

\noi while the pair correlation function $h(s)$ approximately obeys the linearized Boltzmann
equation 

\beq
\label{6.2e}
h(s) = - \beta q \ \phi (s)
\eeq

\noi Thus, we obtain the usual Debye-H\"uckel equation 

\beq
\label{6.3e}
(\Delta - \kappa^2 ) \phi (s) = - 2 \pi q \delta^{(2)} (s)
\eeq

\noi where the inverse Debye length $\kappa$ is defined by

\beq
\label{6.4e}
\kappa^2 = 2 \pi \beta n q^2
\eeq

\noi and here the Laplacian is given by (\ref{3.1e}). In terms of $\tau = s/a$, (\ref{6.3e}) becomes

\beq
\left ( {1 \over \sinh \tau} \ {d \over d \tau} \sinh \tau {d \over d \tau } - \kappa^2 a^2 \right
) \phi = - 2 \pi q \delta^{(2)} (\tau) \label{6.5e}
\eeq
 
\noi The solution which vanishes at infinity is

\beq
\phi = q Q_{\nu} (\cosh \tau )
\label{6.6e}
\eeq

\noi where $Q_{\nu}$ is a Legendre function of the second kind \cite{10r} and 

\beq
\label{6.7e}
\nu = - {1 \over 2} + \left ( {1 \over 4} + \kappa^2 a^2 \right )^{1/2}
\eeq

\noi such that $\nu (\nu +1) = \kappa^2 a^2$. The pair correlation function is given by
(\ref{6.2e}) and the charge pair correlation function is

\beq
\label{6.8e}
\rho^{(2)}(s) = q^2 n^2 h(s) = - {1 \over \beta } \left ( {\kappa^2 \over 2 \pi} \right )^2 Q_{\nu}
(\cosh \tau ) \eeq

\noindent At large distances $s$, (\ref{6.8e}) has the exponential decay $- \exp [- (\nu + 1)s/a]$.
\par

 It can be checked that the screening sum rules (\ref{5.1e}) 	and (\ref{5.3e}) are
satisfied. Indeed, (\ref{5.1e}) results from \cite{10r} 

\beq
\int_1^{\infty} Q_{\nu} (x) dx = {1 \over \nu (\nu + 1)}
\label{6.9}
\eeq

\noi As to (\ref{5.3e}), one can use the Legendre equation (\ref{6.5e}) for showing that $[\nu (\nu
+ 1)]^{-1} (x^2 - 1) dQ_{\nu}(x)/dx$ is a primitive of $Q_{\nu}(x)$, and by successive integrations
by parts of the left-hand side of (\ref{5.3e}) one reduces it to (\ref{6.9}). 

Thus, like in the
case of a flat system, the linearized Debye-H\"uckel approximation exactly obeys the sum rules. This
means that the system is a conductor.

 \subsection{Thermodynamics}
\hspace*{\parindent}
From the correlation function (\ref{6.8e}), one can obtain all the thermodynamic functions, which
are reasonably defined without boundary effects as follows.

The excess internal energy per particle $u_{exc}$ (i.e. the potential energy per particle) is half
the energy of a particle in the potential created by the {\it other} ones~:

\beq
\label{6.9e}
u_{exc} = {1 \over 2} q \lim_{s \to 0} \left [ \phi (s) + q \log \tanh {s \over 2a} \right ]
\eeq

\noi Using the behavior \cite{10r} of $Q_{\nu}$ $(\cosh \tau )$ when $z = \cosh \tau = \cosh (s/a)$
is close to 1,

\beq
\label{6.10e}
Q_{\nu}(z) \sim - {1 \over 2} \log {z - 1 \over 2} - \gamma - \psi (\nu + 1)
\eeq

\noi one finds

\beq
\label{6.11e}
u_{exc} = - {q^2 \over 2} \left [ \gamma + \psi (\nu + 1) \right ]
\eeq

>From (\ref{6.11e}), where $\nu$ is a function of $\beta$ and $n$ through (\ref{6.7e}) and
(\ref{6.4e}), one can obtain the other thermodynamic quantities of interest as defined by
(\ref{4.7}). However, the integral leading to $f$ cannot be expressed in terms of known functions.

 \subsection{Limiting cases}
\hspace*{\parindent}
It can be checked that, in the flat system limit $a \to \infty$, the above results lead to the
known behaviors \cite{13r}

\bea
&&h(x) \sim - \beta q^2 K_0 (\kappa r) \quad , \quad u_{exc} \sim - {q^2 \over 2} \left (
\gamma + \log \kappa a \right ) \quad , \nn \\ 
&&f_{exc} \sim - {q^2 \over 2} \left ( \gamma + \log \kappa a - {1 \over 2} \right ) \quad , \quad
p_{exc} \sim - {q^2 \over 4} n
\label{6.12e}
\eea

The other extreme case is when the density $n$ becomes small for a given value of the curvature
radius $a$. The excess thermodynamic functions are analytical in the density. The excess internal
energy (\ref{6.11e}) has its density expansion starting as

\beq
\label{6.13e}
u_{exc} = - {\pi^3 \over 6} \beta q^4 a^2 n + \cdots
\eeq

\noi in agreement with the weak-coupling limit $\beta q^2 \ll 1$ of (\ref{4.10}). The other
weak-coupling limits are

\beq
\label{6.14e}
f_{exc} = - {\pi^3 \over 12} \beta q^4 a^2 n + \cdots 
\eeq

\beq
\beta p_{exc} = - {\pi^3 \over 12} (\beta q^2)^2 a^2 n^2 + \cdots
\label{6.15e}
\eeq

\mysection{ONE-COMPONENT PLASMA AT $\beta q^2 = 2$}
\hspace*{\parindent}
This is an exactly solvable model in a variety of geometries. On a pseudosphere, the pair correlation
function has been recently obtained by a mapping on a field theory \cite{14r}. Here, all correlation
functions will be derived by adapting the methods which have been used for the previously studied
geometries \cite{15r}.

 \subsection{Correlation functions}
\hspace*{\parindent}
As explained in Section 4.3, one expects to find the same correlation functions either with the
interaction $- \log \tanh (s/2a)$ or with the interaction $- \log \sinh (s/2a)$. It turns out that
it is simpler to use the second one, as follows.

For a one-component plasma with a $- \log \sinh (s/2a)$ interaction, the potential energy can be
obtained by adapting (\ref{4.8e}). Since

\beq
\Delta \log {|z - z'| \over 2a} = 2 \pi \delta^{(2)} (z, z')
\label{7.1e}
\eeq

\noi the particle-background interaction energy arising from the first term of (\ref{4.8e}) is
$W(r)$ such that

\beq
\label{7.2e}
\Delta W(r) = 2 \pi q^2 n
\eeq

\noi The solution of (\ref{7.2e}) is, up to an additive constant,

\beq
\label{7.3e}
W(r) = - 2 \pi q^2 n a^2 \log \left ( 1 - {r^2 \over 4a^2} \right )
\eeq

\noi (choosing this $W$ depending only on the distance $r$ to the center of the Poincar\'e disk is
just a matter of convenience). The second term of (\ref{4.8e}) is another contribution to the
one-body potential. The particle-particle interaction comes from the first term of (\ref{4.8e}).
The background-background interaction is a constant. All together, the total potential energy is

\beq
\label{7.4e}
H = - q^2 \sum_{i>j} \log {|z_i - z_j| \over 2a} - {q^2 \over 2} \left ( 4 \pi n a^2 + 1 \right )
\sum_i \log \left ( 1 - {r_i^2 \over 4a^2} \right ) \eeq

\noi up to some additive constant (actually an infinite one in the thermodynamic limit) which is
irrelevant for the calculation of the correlations.

The formalism for dealing with a potential energy of the form (\ref{7.4e}), in the canonical
ensemble, has been previously developed \cite{15r}. When $\beta q^2 = 2$, the Boltzmann factor is

\beq
\label{7.5e}
e^{-\beta H} = C \left | \prod_{i>j} \left ( z_i - z_j \right ) \prod_i \left ( 1 - {r_i^2 \over
4a^2} \right )^{2 \pi n a^2 + {1 \over 2}} \right |^2 \eeq

\noi where $C$ is a constant. The Vandermonde determinant identity

\beq
\label{7.6e}
\prod_{i>j} \left ( z_i - z_j \right ) = \det \left [ \begin{array}{llll} 1 &1 &1 &\cdots \\ z_1
&z_2 &z_3 &\cdots \\ z^2_1 &z_2^2 &z_3^2 &\cdots \\ \cdots &\cdots &\cdots &\cdots \end{array}
\right ]  \eeq

\noi allows to rewrite (\ref{7.5e}) as the squared modulus of a Slater determinant

\beq
\label{7.7e}
e^{-\beta H} = C \left | \det \left \{ \psi_j (z_i) \right \}_{i,j = 1, 2, 3, \cdots} \right |^2
\eeq

\noi where
\beq
\label{7.8e}
\psi_j(z) = \left ( 1 -{r^2 \over 4a^2} \right )^{2 \pi na^2 + {1 \over 2}} \ z^{j-1}
\eeq

\noi Thus, (\ref{7.7e}) has the form of the squared modulus of the wave function of a system of
independent fermions occupying the mutually orthogonal one-particle wave functions $\psi_j (z)$, and
computing the $n$-body densities is a standard problem. The projector on the functional space
spanned by the functions $\psi_j$ is, in the thermodynamic limit,

\beq
\label{7.9e}
<z_1|P|z_2> = \sum_{j=1}^{\infty} {\psi_j (z_1) \overline{\psi_j (z_2)} \over \int |\psi_j(z)|^2 dS}
\eeq

\noi In terms of this projector, the $n$-particle truncated densities are

\bea
\label{7.10e}
&&n(z) = <z|P|z> \nn \\
&&n_T^{(2)}(z_1, z_2) = - \left | <z_1 |P|z_2> \right |^2 \\
&&n_T^{(n)} (z_1, z_2, \cdots, z_n) = (-)^{n+1} \sum_{(i,i_2 \cdots i_n)} <z_{i_1}|P|z_{i_2}> \cdots
<z_{i_n}|P|z_{i_1}> \nn
\eea

\noi where the summation runs over all cycles ($i, i_2 \cdots i_n$) built with $\{1, 2, \cdots ,
n\}$.

Using (\ref{2.5e}) and (\ref{7.8e}), one computes the normalization integral 

\beq
\label{7.11e}
\int |\psi_j(z)|^2 dS = \pi (4a^2)^j {\Gamma (4 \pi n a^2) \Gamma (j) \over \Gamma (4 \pi n a^2 +
j)}
\eeq

\noi and the series (\ref{7.9e}) can be summed into

\beq
\label{7.12e}
<z_1|P|z_2> = n {\left ( 1 - {r_1^2 \over 4a^2} \right )^{2 \pi na^2 + {1 \over 2}} \left ( 1 -
{r_2^2 \over 4a^2} \right )^{2 \pi na^2 + {1 \over 2}} \over \left ( 1 - {z_1 \bar{z}_2 \over 4a^2}
\right )^{4 \pi n a^2 + 1}}
\eeq

\noi Using (\ref{7.12e}) in (\ref{7.10e}) gives the $n$-body densities.

The one-body density is found to be position-independent, with the value $n$ such that the particle
charge density $qn$ is the opposite of the background charge density, as expected.

After simple manipulations using (\ref{2.4e}), the two-body truncated density (\ref{7.10e}) is
found to be, in agreement with ref. \cite{14r},

\beq
\label{7.13e}
n^2 h(s_{12}) = n_T^{(2)}(z_1, z_2) = - {n^2 \over \left [ \cosh (s_{12}/2a) \right ]^{8 \pi na^2
+ 2}}
\eeq

\noi where $s_{12}$ is the geodesic distance between the two particles. This formula can be
recovered from the analogous result on a sphere \cite{1r} of radius $R$ by replacing $R$ by $ia$.
For large values of the distance $s_{12}$, (\ref{7.13e}) has an exponential decay.

It can be easily checked that the screening sum rules (\ref{5.1e}) and (\ref{5.3e}) are satisfied
by (\ref{7.13e}) (for calculating the integrals, it is convenient to use $\cosh (s/2a)$ as the
variable). Thus, the system is a conductor.

In the Appendix, it is checked that using the interaction $- \log \tanh (s/2a)$ gives the same
correlation functions.

 \subsection{Magnetic analogy}
\hspace*{\parindent}
As in the case of a flat system, there is a magnetic analogy. For a quantum particle of unit charge
living on the surface and submitted to a magnetic field $B$ normal to the surface \cite{16r}, the
wave functions associated with the infinitely degenerate lowest Landau level are of the form
(\ref{7.8e}), in an appropriate gauge, with $2 \pi na^2 + (1/2) = Ba^2$. For a system of
independent fermions filling the lowest Landau level, the two-body truncated density is just
(\ref{7.13e}).

\subsection{Thermodynamics}
\hspace*{\parindent}
The excess internal energy per particle is

\beq
\label{7.14e}
u_{exc} = {n \over 2} \int h(s) q^2v(s) dS
\eeq

\noi where $h(s)$ is given by (\ref{7.13e}), $v(s)$ by (\ref{3.3e}), and $dS$ by (\ref{4.3e}).
Using the variable $t = \tanh^2(s/2a)$, after an integration by parts one obtains

\beq
\label{7.15e}
u_{exc} = {q^2 \over 4} \int_0^1 \left [ (1 - t)^{4 \pi na^2} - 1 \right ] {dt \over t}
\eeq

\noi This is related to an integral representation \cite{10r} of the $\psi$ function such that

\beq
\label{7.16e}
u_{exc} = - {q^2 \over 4} \left [ \gamma + \psi (4 \pi na^2 + 1 ) \right ]
\eeq

In the $a \to \infty$ limit, one correctly recovers the flat system result

\beq
\label{7.17e}
u_{exc} \sim - {q^2 \over 4} \left ( \gamma + \log 4 \pi na^2 \right )
\eeq

\noi The other extreme case is the small-$n$ expansion (indeed, $u_{exc}$ is an analytic function
of $n$ around $n = 0$)

\[
u_{exc} = - {\pi^3 \over 6} q^2 a^2 n + \cdots
\]

\noi which agrees with the result (\ref{4.10}) from the virial expansion with $\beta q^2 = 2$.

Since $u_{exc}$ is known only at one temperature ($\beta q^2 = 2$), it is not possible to use
(\ref{4.7}) for obtaining the free energy per particle and the pressure. Only the beginning of
the virial expansion is available through (\ref{4.6}) and (\ref{4.8})~:

\beq
\label{7.18e}
f_{exc} = - \pi a^2 q^2 n + \cdots
\eeq

\beq
\label{7.19e}
\beta p = n - \pi \beta q^2 a^2 n^2 + \cdots
\eeq

\mysection{TWO-COMPONENT PLASMA AT $\beta q^2 = 2$}
\hspace*{\parindent}
This is a system of two species of particles with charges $q$ and $-q$. When $\beta q^2 = 2$, this
model is exactly solvable in a variety of geometries. In the present geometry of a pseudosphere, the
correlations can be calculated, in the grand-canonical ensemble, by adapting the previously used
methods \cite{2r,17r}.

  \subsection{Correlations} \hspace*{\parindent}
The correlations can be computed by noting that (\ref{4.7e}) or (\ref{4.8e}) map the system onto
a flat system in the Poincar\'e disk. The last term of (\ref{4.7e}) or (\ref{4.8e}) gives to the
Boltzmann factor $\exp (- \beta H)$ a multiplicative contribution $[1 - (r_i/2a)^2]$ for each
particle. Furthermore, in the computation of a partition function, the area element $dS_i$ can be
expressed in terms of the area element $d^2z_i$ as $dS_i = [1 - (r_i/2a)^2]^{-2} d^2z_i$. Thus, the
original system with a constant fugacity $\zeta$ maps onto a flat system with a position-dependent
fugacity $\zeta [1 - (r/2a)^2]^{-1}$.

For that flat system, it results from previous work \cite{17r} that the $n$-body densities can be
expressed in terms of Green functions $\widetilde{G}_{\varepsilon_1 \varepsilon_{2}}(z_1, z_2)$
($\varepsilon_i$ is the sign $\pm$ of the particle at $z_i$). For instance, the one-body density in
the flat disk for particles of sign $\varepsilon$ at $z$ is $4\pi \zeta a [1 - (r/2a)^2]^{-1}$
$\widetilde{G}_{\varepsilon \varepsilon}(z, z)$. The corresponding density on the pseudosphere has a
supplementary multiplicative factor $d^2z/dS$ and is $n_{\varepsilon} = 4 \pi \zeta a$\break
\noindent $[ 1 - (r/2a)^2 ] G_{\varepsilon \varepsilon}(z,z)$.

Thus, in terms of 

\beq
\label{8.1}
G_{\varepsilon_1 \varepsilon_2}(z_1,z_2) = \left ( 1 - {r_1^2 \over 4a^2} \right )^{1/2} \
\widetilde{G}_{\varepsilon_1 \varepsilon_2}(z_1, z_2) \left ( 1 - {r_2^2 \over 4a^2} \right )^{1/2}
 \eeq

\noi the one-body densities are

\bminiG{EDh}
\label{8.1ae}
n_{\varepsilon} = 4 \pi \zeta a G_{\varepsilon \varepsilon}(z,z)
\eeeq
\noi Similarly, the two-body truncated densities are 
\beeq
\label{8.1be}
n_{\varepsilon_1 \varepsilon_2}^{(2)T}(z_1, z_2) = - \varepsilon_1 \varepsilon_2 (4 \pi \zeta a)^2
  \left | G_{\varepsilon_1
\varepsilon_2}(z_1, z_2) \right |^2
 \emini

\noi and higher-order correlations are given by sums over cycles similar to the ones in
(\ref{7.10e}), the elements of the cycles now being $4 \pi \zeta a G_{\varepsilon_i
\varepsilon_j}(z_i, z_j)$. 

The Green functions $\widetilde{G}_{\varepsilon \varepsilon '} (z, z')$ form a $2 \times 2$ matrix
$\widetilde{G}$. It was shown \cite{17r} that, when the potential energy is of the form (\ref{4.8e}),
$\widetilde{G}$ obeys the equation

\beq
\label{8.2e}
\left [ {\partial \hskip - 2.5 truemm /} + {4 \pi \zeta a \over 1 - (r/2a)^2} \right ]
\widetilde{G}(z, z') = {\bf I} \ \delta^{(2)}_{flat}(z, z') \eeq

\noi where ${\partial \hskip - 2.5 truemm /}$ is the flat Dirac operator

\beq
\label{8.3e}
{\partial \hskip - 2.5 truemm /} = 2 \left ( \begin{array}{cc} 0 & \partial_z \cr \partial_{\bar{z}}
&0 \cr \end{array} \right ) \eeq

\noi and ${\bf I}$ the unit $2 \times 2$ matrix~; $\delta_{flat}^{(2)}$ is the usual Dirac
distribution in the plane. From (\ref{8.1}) and (\ref{8.2e}), one finds

\beq
\label{8.5}
( {D \hskip - 2.5 truemm /} + 4 \pi \zeta a ) \ G(z, z') = {\bf I} \ \delta^{(2)}(z, z') 
\eeq
 
\noi where

\beq
\label{8.6}
{D \hskip - 2.5 truemm /} = \left ( 1 - {r^2 \over 4a^2} \right )^{3/2} \ {\partial \hskip - 2.5 truemm
/} \left ( 1 - {r^2 \over 4a^2} \right )^{-1/2}  \eeq

\noi is the Dirac operator on the pseudosphere, and now $\delta^{(2)}$ is the Dirac distribution on
the pseudosphere, obeying (\ref{3.3}) . Thus $G$ is the Green function of ${D \hskip - 2.5 truemm
/} + 4 \pi \zeta a$, on the pseudosphere, and the densities have the simple expressions (8.2) in
terms of $G$.

The operator $(1/4 \pi a){\partial \hskip - 2.5 truemm /}$ appeared in
the formalism as the inverse of a matrix kernel

\beq
\label{8.4e}
M(z, z') = \left ( \begin{array}{cc} 0 &{2a \over z - z'} \cr {2a \over \bar{z} - \bar{z}'} &0 \cr
\end{array} \right ) 
\eeq

\noi associated to the potential energy (\ref{4.8e})~: 

\beq
\label{8.5e}
{1 \over 4 \pi a} \ {\partial \hskip - 2.5 truemm /} M(z, z') = {\bf I} \  \delta_{flat}^{(2)} (z,
z') \eeq

\noi If, instead of (\ref{4.8e}), one uses the potential energy (\ref{4.7e}) which is like the one of
a system with ideal conductor walls, (\ref{8.4e}) is replaced by \cite{18r,19r}

\beq
\label{8.6e}
M(z, z') = \left ( \begin{array}{cc} {4a^2 \over 4a^2 - z \bar{z}'} &{2a \over z - z'} \cr {2a
\over \bar{z} - \bar{z}'} &{4a^2 \over 4a^2 - \bar{z} z'} \cr \end{array} \right )
\eeq

\noi However, (\ref{8.5e}) remains valid (because $4a^2 - z\bar{z}'$ has no zero inside the
Poincar\'e disk), and one finds for $\widetilde{G}$ the same equation (\ref{8.2e}). It can be shown
that in both cases (\ref{4.7e}) and (\ref{4.8e}), $\widetilde{G}(z, z')$ should go to zero at
infinity, i.e. when $z$ goes to the boundary circle of the Poincar\'e disk. Therefore, the
correlations are the same ones for a two-body interaction $\mp q^2 \log \sinh (s/2a)$ or $\mp q^2
\log \tanh (s/2a)$.

One element of the matrix equation (\ref{8.2e}) is 

\beq
\label{8.7e}
2 \partial_{\bar{z}} G_{++}(z, z') + {4 \pi \zeta a \over 1 - (r/2a)^2} \  G_{-+} (z, z') = 0
\eeq

\noi By combining (\ref{8.7e}) with the other equation relating $\widetilde{G}_{++}$ and
$\widetilde{G}_{-+}$, one finds an equation for $\widetilde{G}_{++}$ alone~:

\beq
\label{8.8e}
\left [ - 4 \partial_z \left ( 1 - {r^2 \over 4a^2} \right ) \partial_{\bar{z}} + {(4 \pi \zeta
a)^2 \over 1 - (r/2a)^2} \right ] \widetilde{G}_{++}(z, z') = 4 \pi \zeta a \delta^{(2)}_{flat} (z,
z')   \eeq

\noi Similar equations hold for $\widetilde{G}_{--}$ and $\widetilde{G}_{+-}$. Since
$n_{\varepsilon}$ is expected to be position-independent and $n_{\varepsilon_1 
\varepsilon_2}^{(2)T}(z_1, z_2)$ to depend only on the geodesic distance $s$ between points $z_1$
and $z_2$, for obtaining these quantities it is enough to compute $\widetilde{G}(z, z')$ in the
simplest case $z' = 0$. One looks for a solution of (\ref{8.8e}) depending only on $r$ and one uses
the variable $u = \cosh^2(s/2a) = [1 - (r/2a)^2]^{-1}$. Then, (\ref{8.8e}) becomes the
hypergeometric equation

\beq
\label{8.9e}
\left [ u (1 - u) {d^2 \over du^2} - u {d \over du} + (4 \pi \zeta a^2)^2 \right ]
\widetilde{G}_{++} = 0 \quad , \quad u > 1 \eeq

\noi with the boundary condition that $\widetilde{G}_{++}$ behaves like $- \zeta a \log (u - 1)$ as
$r \to 0$ ($u \to 1$). Furthermore, $\widetilde{G}_{++}$ should vanish at infinity ($u \to \infty$).
The solution of (\ref{8.9e}) which satisfies these conditions is \cite{10r}

\beq
\label{8.10e}
\widetilde{G}_{++} (z, 0) = \zeta a {\Gamma (4 \pi \zeta a^2) \Gamma (4 \pi \zeta a^2 + 1) \over
\Gamma (8 \pi \zeta a^2 + 1)} u^{- 4 \pi \zeta a^2} F \left ( 4 \pi \zeta a^2, 4 \pi \zeta a^2 + 1 ;
8 \pi \zeta a^2 + 1 ; u^{-1} \right )
\eeq

\noi Using (\ref{8.7e}) gives \cite{10r}

\bea
\label{8.11e}
\widetilde{G}_{-+} (z, 0) &=& e^{i \varphi} \zeta a {\Gamma (4 \pi \zeta a^2) \Gamma (4 \pi \zeta
a^2 + 1) \over \Gamma (8 \pi \zeta a^2 + 1)} (1 - u^{- 1})^{1/2} u^{-4 \pi \zeta a^2} \nn \\ 
&&\times F \left ( 4 \pi \zeta a^2 + 1, 4 
\pi \zeta a^2 + 1 ; 8 \pi \zeta a^2 + 1 ; u^{-1} \right ) 
\eea 

\noi One also finds $G_{--}(z, 0) = G_{++}(z, 0)$ and $G_{+-}(z, 0) = - \overline{G_{-+}(z, 0)}$.
By using the expansions of the hypergeometric functions with respect to $1 - u^{-1}$, one can check
that these Green functions have the proper flat-system limits as $a \to \infty$ for a fixed value
of $m = 4 \pi a \zeta$.

Using these results in (\ref{8.1}) and (8.2) gives the one- and two-body densities.
However, $G_{\varepsilon \varepsilon}(z, 0)$ has a logarithmic divergence as $z \to 0$, and one
should use $G_{\varepsilon \varepsilon} (\sigma ,  0)$ in (\ref{8.1ae}), where $\sigma$ is a small
cutoff, rather than $G_{\varepsilon \varepsilon} (0, 0)$. This amounts to replacing the point
particles by small hard disks of diameter $\sigma$. The final results are

\bminiG{EDh}
\label{8.12ae}
n_{\pm} = \zeta (4 \pi \zeta a^2) \left [ - 2 \gamma - \psi (4 \pi \zeta a^2) - \psi (4 \pi \zeta
a^2 + 1) + 2 \log {2a \over \sigma} \right ] \eeeq
\noi (the expansion \cite{10r} of (\ref{8.10e}) with respect to $1 - u^{-1}$ has been used),  
\beeq
\label{8.12be}
&&n_{++}^{(2)T}(s) = n_{--}^{(2)T}(s) = - \zeta^2 {[\Gamma(4 \pi \zeta a^2 + 1)]^4 \over [ \Gamma
(8 \pi \zeta a^2 + 1)]^2} {1 \over [ \cosh (s/2a)]^{16 \pi \zeta a^2 + 2}} \nn \\
&&\times \left [ F \left ( 4 \pi \zeta a^2, 4 \pi \zeta a^2 + 1; 8 \pi \zeta a^2 + 1 ; {1 \over (
\cosh {s \over 2a})^2} \right ) \right ]^2 \eeeq 
\beeq
\label{8.12ce}
&&n_{-+}^{(2)T}(s) = n_{+-}^{(2)T}(s) =  \zeta^2 {[\Gamma(4 \pi \zeta a^2 + 1)]^4 \over [ \Gamma
(8 \pi \zeta a^2 + 1)]^2} {[\tanh (s/2a)]^2 \over [ \cosh (s/2a)]^{16 \pi \zeta a^2 + 2}} \nn \\
&&\times \left [ F \left ( 4 \pi \zeta a^2 + 1, 4 \pi \zeta a^2 + 1; 8 \pi \zeta a^2 + 1 ; {1 \over (
\cosh {s \over 2a})^2} \right ) \right ]^2
 \emini

For the calculation of higher-order $n$-body densities, the Green functions $G(z, z')$ with
arbitrary $z'$ are needed. They would be obtained by a conformal transformation such that 0 maps
onto $z'$, in the same way as in the case of a sphere \cite{2r}.

At small distances $s$, (\ref{8.12be}) and (\ref{8.12ce}) behave like in the flat system case
\cite{17r}, i.e. as $- (\log s)^2$ and $1/s^2$, respectively. At large distances, they both have an
exponential decay as $\mp \exp [-(8 \pi \zeta a^2 + 1)s/a]$. 

The strong-curvature limit $a \to 0$ is also a low-density limit $\zeta a^2 \to 0$. In that limit
(\ref{8.12be}) and (\ref{8.12ce}) should behave like $\mp \zeta^2 \beta q^2 v \sim \mp 4 \zeta^2
\exp (- s/a)$ and they do. 

\subsection{Thermodynamics} \hspace*{\parindent}
Since the total density $n = 2n_+$ is given as a function of the fugacity $\zeta$ by
(\ref{8.12ae}), the pressure can be obtained by integrating $n = \zeta d(\beta p)/d \zeta$.
However, the integration cannot be performed in terms of known functions for arbitrary $\zeta$.
Explicit results can be obtained in limiting cases only.

The flat system results \cite{17r} can be recovered as $a \to \infty$ for a fixed value of $m = 4 \pi
a \zeta$.

In the opposite limiting case $\zeta \to 0$, for a given value of $a$, one obtains

\beq
\label{8.16e}
n = 2n_{\pm} = 2 \zeta + \left ( 16 \pi a^2 \log {2a \over \sigma} \right ) \zeta^2 + \cdots
\eeq 

\noi which gives

\beq
\label{8.17e}
\beta p = 2 \zeta + \left ( 8 \pi a^2 \log {2a \over \sigma} \right ) \zeta^2 + \cdots
\eeq 

\noi and therefore

\beq 
\label{8.18e}
\beta p = n - \left ( 2 \pi a^2 \log {2a \over \sigma} \right ) n^2 + \cdots
\eeq

\noi This is in agreement with a direct calculation of the second virial coefficient (\ref{4.2e})
with a lower cutoff on $s$ at $s = \sigma$.

\mysection{CONCLUSION}
\hspace*{\parindent}
The present paper has already been summarized in the Abstract. Here are some final remarks.

Perhaps a possible way of computing a free energy without boundary effects would be to start with a
finite domain of suitable shape with periodic boundary conditions and to take its thermodynamic
limit. However, we have not been able to use this seemingly difficult approach. 

It is likely that the Kosterlitz-Thouless transition \cite{20r} of a two-component plasma to a
dielectric phase at low temperature does not exist on a pseudosphere. Since the potential $- \log
\tanh (s/2a)$ now goes to zero at infinity, the original argument about a balance between energy
and entropy here says that breaking a pair of oppositely charged particles always leads to a lower
free energy because the entropy wins. The two-component plasma is expected to remain a conductor
even at low temperature.\footnote{An opposite case when there is no Kosterlitz-Thouless phase
transition is a two-component plasma on the surface of a cylinder of finite radius and infinite
length \cite{21r}. At large separations, the interaction potential behaves like the distance, the
system behaves like a one-dimensional two-component plasma and is always in a dielectric phase
\protect{\cite{22r}}.}

When $\beta q^2 = 2$, i.e. when the temperature is twice the Kosterlitz-Thouless transition
temperature for a flat system, the two-component plasma is a conductor even in a flat space, and
{\it a fortiori} should be a conductor on a surface of constant negative curvature. It would be
of some interest to check that conjecture by explicitly showing that the modified Stillinger-Lovett
sum rule (\ref{5.3e}) is satisfied by the correlation functions (\ref{8.12be}) and (\ref{8.12ce}).
This is left as an open problem, because of the technical difficulty of computing the relevant
integral.

Another, more fundamental, open problem is how to compute the pressure of the one-component plasma
at the temperature at which the pair correlation function (\ref{7.13e}) is known. For a flat
system, one would use the virial relation

\beq
p = \beta^{-1} n - {n^2 \over 4} q^2 \int h(r) {dv \over dr} \ r \ dS
\label{9.1e}
\eeq

\noi Is there a generalization of this relation on a pseudosphere~? 

\newpage
\appendix
\mysection{APPENDIX : MORE ON THE CORRELATION FUNCTIONS OF THE ONE-COM\-PO\-NENT PLASMA}
The correlation functions of Section 7.1 can also be obtained when using the in\-te\-rac\-tion $-
\log \tanh (s/2a)$, by adapting the grand canonical formalism used in Section 8.1 for the
two-component plasma.

Eq. (\ref{8.2e}) could have been obtained from the integral equation \cite{18r,19r}

\beq
\widetilde{G}(z, z') + \int M(z, z'') {\zeta \over 1 - (r''/2a)^2} \widetilde{G}(z'',z') d^2z'' = {1
\over 4 \pi a} M(z, z')\label{A.1e} \eeq

\noi where $M$ is the matrix (\ref{8.4e}). Indeed, applying ${\partial \hskip - 2.5 truemm /}$ to
both sides of (\ref{A.1e}) gives (\ref{8.2e}) when (\ref{8.5e}) is taken into account.

Eq. (\ref{A.1e}) can be adapted to the case of a one-component plasma by keeping only the $++$
components of the matrices $\widetilde{G}$ and $M$ (from now on, $\widetilde{G}_{++}$ will be called
$\widetilde{G}$). Furthermore, the particle-background interaction (\ref{7.3e}) brings into the
position-dependent fugacity a supplementary multiplicative factor

\beq
\label{A.2e}
e^{-\beta W} = e^C \left [ 1 - \left ( {r \over 2a} \right )^2 \right ]^{4 \pi na^2}
\eeq

\noi where $C$ comes from the additive constant in the particle-background interaction. Here $C$ is
important, having the property $C \to + \infty$ in the thermodynamic limit. With these
modifications, (\ref{A.1e}) becomes

\beq
\label{A.3e}
\widetilde{G}(z,z') + \zeta e^C \int {1 \over 1 - (z \bar{z}''/4a^2)} \left ( 1 - {r''^2 \over 4a^2}
\right ) ^{4 \pi na^2 - 1} \widetilde{G}(z'',z') d^2z'' 
= {1 \over 4 \pi a \left [ 1 - (z \bar{z}'/4a^2) \right ]}  \eeq

\noi From (\ref{A.3e}), it results that $\widetilde{G}(z, z')$ is an analytic function of $z$. For
simplicity, we shall consider only the case $z' = 0$. If one looks for a solution of circular
symmetry, $\widetilde{G}(z, 0)$ is necessarily a constant, and (\ref{A.3e}) becomes

\beq
\label{A.4e}
\widetilde{G} \left [ 1 + {\zeta e^C \over n} \right ] = {1 \over 4 \pi a}
\eeq

\noi The $n$-body densities involve $4 \pi a \zeta e^C \widetilde{G}$ which, in the thermodynamic
limit $C \to \infty$, becomes the background density $n$, independent of the fugacity $\zeta$.
Including in (\ref{8.1ae}) and (\ref{8.1be}) the multiplicative factor (\ref{A.2e}) for each
fugacity gives $n$ for the particle density, and

\beq
\label{A.5e}
n^{(2)}_T(z,0) = - n^2 \left [ 1 - \left ( {r \over 2a} \right )^2 \right ]^{4 \pi na^2}
\eeq 

\noi for the two-body truncated density in agreement with (\ref{7.13e}).

The above calculation illustrates that, in a grand-canonical approach to the one-component plasma
with a fixed background, the bulk properties are fugacity-independent. \par \vskip 5 truemm

\section*{ACKNOWLEDGEMENTS} \par
We thank M. B. Hastings for stimulating e-mail exchanges. We also thank A. Comtet for having taught
us the facts of life about spaces of constant negative curvature, and for his help in developing
the magnetic analogy briefly discussed in Section 7.2.

We give credit to the referee who made the remark added to the Conclusion as a footnote.


\newpage
\begin{thebibliography}{99} \par \vskip 5 truemm  

\bibitem{1r} J. M. Caillol, {\it J. Physique - Lettres} {\bf 42}:L-245 (1981).  

\bibitem{2r} P. J. Forrester, B. Jancovici, and J. Madore, {\it J. Stat. Phys.} {\bf 69}:179
(1992).  
  
\bibitem{3r} P. J. Forrester and B. Jancovici, {\it J. Stat. Phys.} {\bf 84}:337 (1996). 
  
\bibitem{4r} J. M. Caillol and D. Levesque, {\it J. Chem. Phys.} {\bf 94}:597 (1991). 
  
\bibitem{5r} J. M. Caillol, {\it J. Chem. Phys.} {\bf 99}:8953 (1993). 

\bibitem{6r} See e.g., N. L. Balazs and A. Voros, {\it Phys. Rep.} {\bf 143}:109 (1986). 

  \bibitem{7r} G. Manificat and J. M. Caillol, {\it J. Chem. Phys.} {\bf 103}:4266 (1995).
 
\bibitem{8r} G. Gallavotti and F. Nicol\'o, {\it J. Stat. Phys.} {\bf 39}:133 (1985). 

\bibitem{9r} P. J. Forrester, {\it J. Stat. Phys.} {\bf 60}:203 (1990). 

 \bibitem{10r} A. Erd\'elyi et al., {\it Higher Transcendental Functions} (Mc Graw-Hill, New York,
1953), Vol. I. 

\bibitem{11r} F. H. Stillinger and R. Lovett, {\it J. Chem. Phys.} {\bf 49}:1991 (1968). 

\bibitem{12r} S. L. Carnie and D. Y. C. Chan, {\it Chem. Phys. Lett.} {\bf 77}:437 (1981). 

\bibitem{13r} C. Deutsch, H. E. DeWitt, and Y. Furutani, {\it Phys. Rev. A} {\bf 20}:2631 (1979). 

\bibitem{14r} M. B. Hastings, {\it J. Stat. Phys.}, to be published. 

  \bibitem{15r} F. Cornu, B. Jancovici, and L. Blum, {\it J. Stat. Phys.} {\bf 50}:1221 (1988). 
  
\bibitem{16r} A. Comtet, {\it Ann. Phys.} {\bf 173}:185 (1987). 

\bibitem{17r} F. Cornu and B. Jancovici, {\it J. Chem. Phys.} {\bf 90}:2444 (1989).
 
\bibitem{18r} P. J. Forrester, {\it J. Chem. Phys.} {\bf 95}:4545 (1991). 

\bibitem{19r} B. Jancovici and G. T\'ellez, {\it J. Stat. Phys.} {\bf 82}:609 (1996).

\bibitem{20r} J. M. Kosterlitz and D. J. Thouless, {\it J. Phys. C} {\bf 6}:1181 (1973).

\bibitem{21r} P. J. Forrester, {\it J. Stat. Phys.} {\bf 63}:491 (1991).

\bibitem{22r} M. Aizenman and P. A. Martin, {\it Commun. Math. Phys.} {\bf 78}:99 (1980). 
\end{thebibliography}
\end{document}